\documentclass[prl,aps,twocolumn,superscriptaddress]{revtex4-1}
\usepackage{latexsym,amssymb,amsfonts,amsmath,graphicx,bbm,mathptm,dsfont,color}

\usepackage{graphicx}

\def\tr{\mbox{tr}}
\def\bra#1{\langle{#1}|}
\def\ket#1{|{#1}\rangle}
\def\braket#1{\langle{#1}\rangle}

{\catcode`\|=\active 
  \gdef\Braket#1{\begingroup
\mathcode`\|32768\let|\BraVert\left<{#1}\right>\endgroup}}
\def\BraVert{\egroup\,\mid\,\bgroup}

\newcommand{\f}{\mathbf{F}}
\newcommand{\s}{\mathbf{S}}
\newcommand{\U}{\mathbf{U}}
\newcommand{\w}{\mathbf{W}}
\newcommand{\q}{\mathbf{Q}}
\newcommand{\x}{\mathbf{X}}

\definecolor{Red}{rgb}{1,0,0}
\definecolor{Blue}{rgb}{0,0,1}
\allowdisplaybreaks

\begin{document}

\title{Energetic fluctuations in an open quantum process}

\author{John Goold}
\affiliation{The Abdus Salam International Centre for Theoretical Physics (ICTP), Trieste, Italy}

\author{Kavan Modi}
\affiliation{School of Physics, Monash University, VIC 3800, Australia}

\date{\today}

\begin{abstract}
Relations similar to work and exchange fluctuations have been recently derived for open systems dynamically evolving in the presence of an ancilla. Extending these relations and constructing a non-equilibrium Helmholtz equation we derive a general expression for the energetic and entropic changes of an open quantum system undergoing a nontrivial evolution. The expressions depend only on the state of the system and the dynamical map generating the evolution. Furthermore our formalism makes no assumption on either the nature or dimension of the ancilla. Our results are expected to find application in understanding the energetics of complex quantum systems undergoing open dynamics.
\end{abstract}

\maketitle

%*****************************************
%*****************************************
{\bf Introduction.---}
Entropy production is instrumental in the analysis of many non-equilibrium effects in different branches of physics and is key to the proper evaluation of the efficiency and performance of thermodynamic devices~\cite{groot}. The thermodynamics of general open quantum systems is a mature and ongoing research program. Many results have been obtained in the case of weakly coupled and slowly driven systems where the dynamics is well described by Markovian master equations~\cite{lindblad, breurbook, alicki, lindblad2} and in particular microscopic expressions for the entropy production are known~\cite{spohn}.

One way to describe the thermodynamics of systems where thermodynamic (and quantum) fluctuations cannot be ignored is by using the work and exchange fluctuation relations which have been demonstrated in the classical domain experimentally~\cite{jrev, Jarzynski:97, jarzynski:04, Crooks}. The fluctuation relations, extended to the quantum mechanical domain~\cite{Tasaki, mrev} are a promising route to understand the thermodynamics of small quantum systems which are operating under non-equilibrium conditions. Recent work has demonstrated that the fluctuation formalism is a tangible route for the experimental exploration of quantum thermodynamics~\cite{dorner2, mauro, fermi, nmr,heat}.

With the surge of interest in the thermodynamics of quantum systems and the development of quantum fluctuation relations research has been directed to microscopic expressions for entropy production~\cite{deffner, cvdb}. Nevertheless extension of the fluctuation formalism to the open quantum system framework leads to some difficulties without making some fairly restrictive assumptions~\cite{esposito, lindblad}. A recent series of papers have analyzed fluctuation-like relations from the operational viewpoint employing the full machinery of completely positive and trace preserving (CPTP) maps~\cite{Crooks, kafri, vedral, rastegin, albash, rastegin2, bound} which are ubiquitous in quantum information. It was found that fluctuation relations, of the standard form, can be derived if the map generated by the open dynamics obeys the unital condition. This has been connected to the breakdown of the ``micro-reversibility" for non-unital quantum channels. At first sight it appears problematic to describe a thermodynamics for the most general type of evolution of a quantum system.

In this work we show how the fluctuation-like relations derived with CPTP maps maybe used in order to connect directly with the thermodynamic laws. Using the relations we show that the energy change of an open quantum system can be divided up into different terms each with its own operational meaning. Furthermore, we construct a non-equilibrium Helmholtz equation and by combining with the fluctuation relations we find a general law for the entropy change of an open system which crucially only depends on the system state and map generating the dynamics.  

%*****************************************
%*****************************************
{\bf Thermodynamic of open processes.---}
When a open system reaches equilibrium it will be in a thermal state at inverse temperature $\beta$. This is the zeroth law of thermodynamics. The free energy of the system and its non-equilibrium entropy are defined using the reservoir as a reference. The fist law of thermodynamics generalised to a non-equilibrium thermodynamic transformation~\cite{deffner} reads
\begin{align}
\label{first}
\Delta \U = \braket{\w} + \braket{\q} \quad \mbox{with} \quad
\braket{\w} = \Delta \f + \braket{\w}_{diss},
\end{align}
here $\Delta \U$ is the internal energy change, $\braket{\w}$ is the average work done and the $\braket{\q}$ is the average heat exchanged, and the average work can be divided into the equilibrium free energy change $\Delta \f$ and the dissipated work $\braket{\w}_{diss}$. The second law of thermodynamics is expressed as
\begin{gather}
\label{second}
\Delta \s = \beta \braket{\q} + \braket{\Sigma},
\end{gather}
where $\Delta \s$ is the change of entropy induced by the thermodynamic transformation and $\braket{\Sigma}$ is the average irreversible entropy produced. The quantities in these laws maybe related to the average of distributions which are typically used 
to formulate various fluctuation relations~\cite{mrev}. 

%%%%%%%%%%%%%%%%%%%%%%%%%%%%%%%%%%%%%%%%%%%%%%%%%
\begin{figure}[t]
\includegraphics[scale=.4]{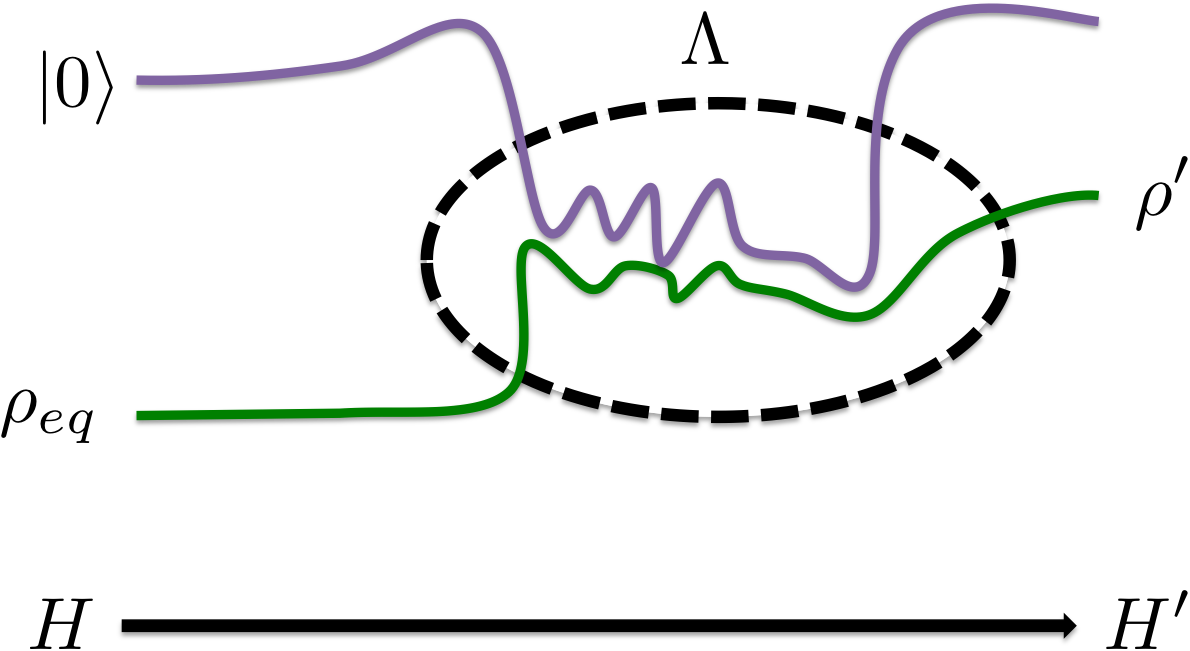}
\caption{\emph{The forward open process.} The system (green line) is initially in a thermal state $\rho_{eq}$ of Hamiltonian $H$. An ancilla (purple line) comes along and exchanges energy with the system. The initial state of the ancilla can always be taken to be $\ket{0}$. However, it can be taken be the maximally mixed state when the map is unital. The whole exchange, shown in the dotted oval, is described by the map $\Lambda$. While the system is interacting with the ancilla, its Hamiltonian is also changed to $H'$. The final state of the system is $\rho'$. \label{fig1}}
\end{figure}
%%%%%%%%%%%%%%%%%%%%%%%%%%%%%%%%%%%%%%%%%%%%%%%%%

Suppose our system is isolated and initially in equilibrium with the reservoir at inverse temperature $\beta$. Then an ancilla comes along and exchanges energy with the system, this is illustrated in Fig.~\ref{fig1}. In general, some work may be performed on the system as well some heat may be dissipated  during this interaction. This is the generation of the open dynamics for the mesoscopic system and we stress that we are not fixing the Hamiltonian of our system--it may change and therefore both heat and work processes occur in the system. In fact, for now, we give up the ability to differentiate between them and will only talk about the average change in internal energy of the system induced by the overall open thermodynamic process. As is customary in the derivation of fluctuation relations let us define initial and final equilibrium states.
\begin{align}
& \rho_{eq} = \frac{1}{Z} \sum_m e^{- \beta E_m} \ket{E_m}\bra{E_m} \quad \mbox{and} \notag\\  
& \rho'_{eq} = \frac{1}{Z'} \sum_m e^{- \beta E'_m} \ket{E'_m}\bra{E'_m}.
\end{align}
Let us assume we have a generic open process which is described by a completely-positive trace-preserving map 
\begin{gather}
\Lambda(\rho) = \sum_l A_l \, \rho \, A^\dag_l = \mbox{tr}_\mathcal{A}[U \rho \otimes \ket{0}\bra{0} U^\dag] = \rho',
\end{gather}
where $A_l = \braket{l | U |0}$. A CPTP maps is trace preserving if $\sum_l A^\dag_l A_l = \openone$ and it is unital ($\Lambda(\openone)=\openone$) if and only if $\sum_l A_l A^\dag_l = \openone$, but we are not assuming the last condition. For a generic map the state of the ancilla can be taken to be a pure state $\ket{0}$~\cite{Nielsen}. However, for unital maps the ancilla can be taken to be the maximally mixed state.

%*****************************************
%*****************************************
{\bf Forward distribution and Jarzynski equality.---}
We can compute a distribution of change in energy for such a CPTP process. The process (which we now call `forward' process for reasons which will become clear) can be any form of thermodynamic process ranging from adiabatic dynamics to highly out of equilibrium dynamics. We are interested in the distribution of internal energy induced by the process which is given by
\begin{align}
P_F(\Delta \U) = \sum_{lmn} & \braket{E'_n | A_l | E_m} \braket{E_m| \rho_{eq} |E_m} \braket{E_m | A^\dag_l | E'_n} \notag \\ & \times \delta(\Delta \U-(E'_n-E_m)).
\end{align}
The distribution comes from three steps: We first measure which energy-eigenstate the system; then we evolve it via the map $\Lambda$; and finally we measure the corresponding distribution of energies of $H'$. The distribution can be shown to be a normalized: $\int d\Delta \U \, P_F(\Delta \U) =1$, see Appendix for details. The first moment of this distribution is the change in internal energy which enters in the first law above: $\Delta \U=\tr(H_{f} \, \rho')-\tr(H_{i} \, \rho_{eq})$. 

It is illustrative to derive a Jarzynski-like equality~\cite{Jarzynski:97} on the distribution of energy changes of the open system in line with \cite{kafri, vedral, rastegin, albash, rastegin2,bound}. We stress that here we are looking at the distribution of internal energy changes and this cannot be associated the ``work" in general as we are allowing for both ``heat" and ``work" process to occur simultaneously. 
We stress that the distribution of energy is expected not to obey a fluctuation relation as pointed out by Talkner {\it et al.} \cite{michelle}. However, as pointed out in \cite{kafri, rastegin, albash, rastegin2,bound} relationships bearing the same mathematical form as fluctuation relations can be derived providing the map has what is known as the unital property (this means that the identity is a fixed point). Interestingly this seemingly simple property has surprising and profound thermodynamic repercussions, see also \cite{binder} for an alternative view.

The derivation of the Jarzynski-like equality on the energy distribution is given in the Appendix and reads:
\begin{align}
\braket{e^{-\beta \Delta \U + \beta \Delta \f}} =& \mbox{tr} \left[ \sum_{l} A_l A^\dag_l \rho'_{eq} \right] := \gamma := e^{-\beta \x}.
\end{align}
From this we can easily see that a Jarzynski-like relation holds, that is, when the channel is unital we have $\sum_{l} A_l A^\dag_l = \openone$ and therefore $\gamma=1$.

\begin{figure}[t]
\includegraphics[scale=.4]{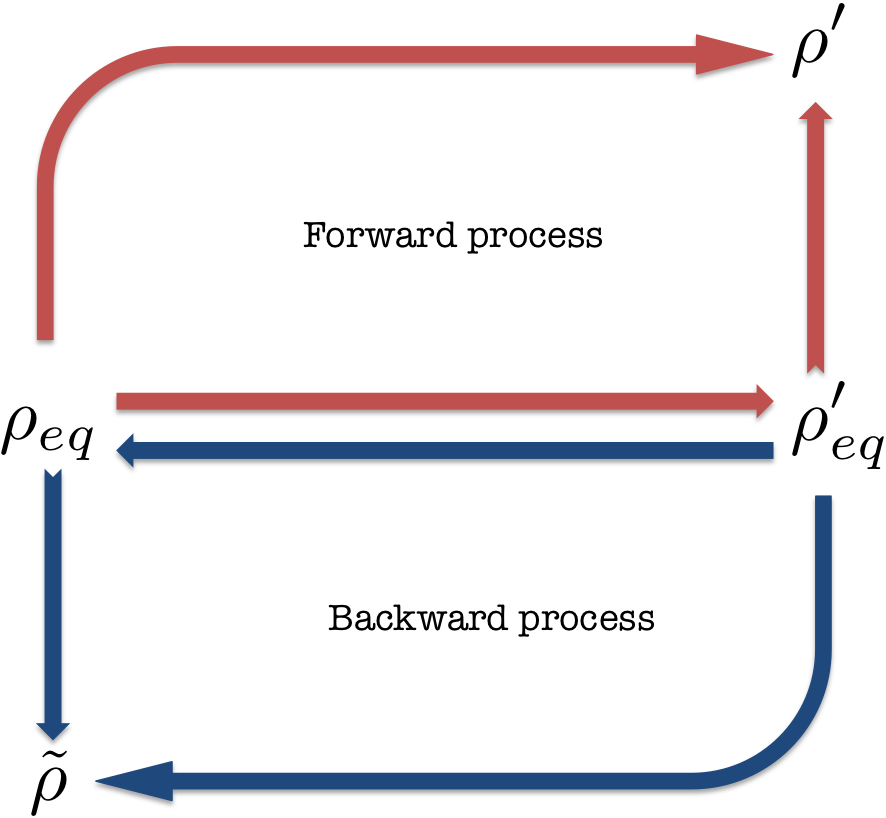}
\caption{\emph{Crooks' setup.} To derive the Crooks' relation we consider the following setup. The forward process, generated by a completely-positive trace-preserving map, is shown on the top (in red). The backward process, generated by a completely-positive but not trace-preserving map, is shown on the bottom (in blue). Both processes are broken up into an isothermal process $\rho_{eq} \to \rho'_{eq}$, and process at a constant Hamiltonian, i.e., $\rho'_{eq} \to \rho'$ at constant $H'$ and $\rho_{eq} \to \tilde{\rho}$ at constant $H$.
\label{fig2}}
\end{figure}

%*****************************************
%*****************************************
{\bf Backward distribution and Crooks' relation.---}
We now turn towards the Crooks relation \cite{Crooks}. For this purpose we need to define a reverse or ``backward" process (see Fig.~\ref{fig2} for an illustration). When the forward process is unital, the backward process can be any trace preserving process. This is because the initial state of the ancilla is can be taken to be maximally mixed, i.e., any possible state and consequently the backward process does not have to yield a specific state of the ancilla at the end of the backward process.

\begin{figure}[t]
\includegraphics[scale=.4]{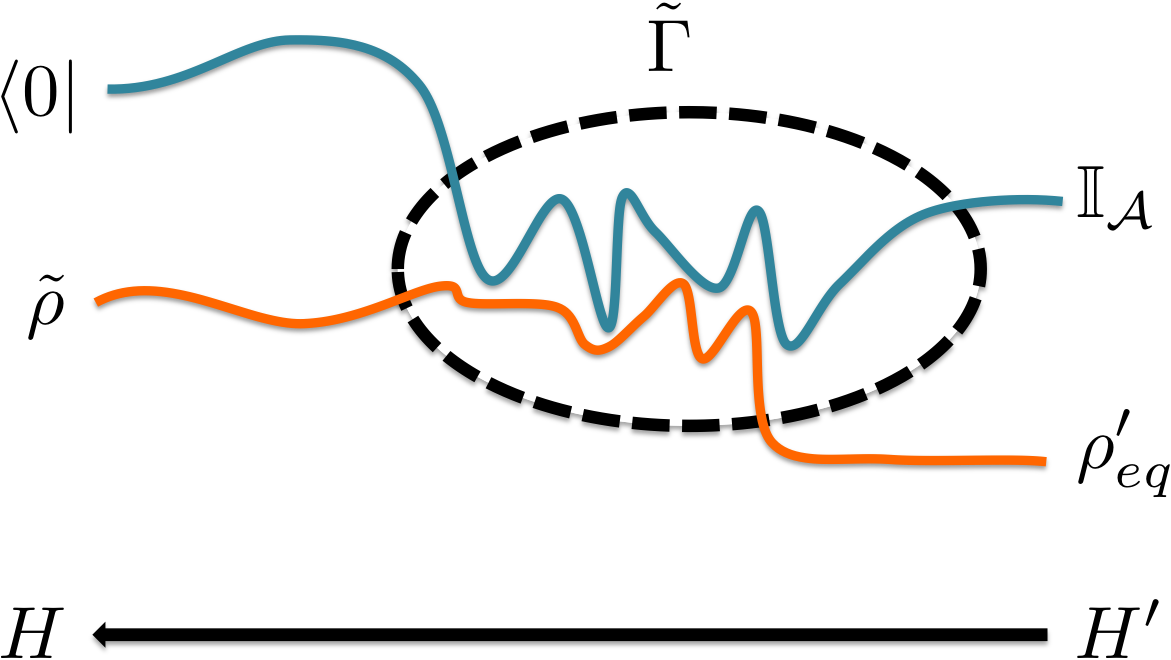}
\caption{\emph{The backward open process.} The system (orange line) is initially in a thermal state $\rho'_{eq}$ of Hamiltonian $H'$. Since the backward process is unital the ancilla (blue line) is allowed to be in any state, which is on average maximally mixed. The system and the ancilla exchanges energy, shown in the dotted oval, is described by the map $\tilde\Gamma$. While the system is interacting with the ancilla, its Hamiltonian is also changed to $H$. The ancilla is then projected onto state $\ket{0}$, which guarantees that the ancilla is back in the state it started at the beginning of the forward process. When this happens the final unnormalised state of the system is $\tilde\rho$.
 \label{fig3}}
\end{figure}

When the forward process is non-unital, the initial state of the ancilla cannot be fully mixed. Therefore, the corresponding backward process must yield the initial state of the ancilla at the end of the backward process. More precisely, we may define the backward process as
\begin{gather}\label{backward}
\tilde\Gamma(\rho') = \sum_l B^\dag_l \, \rho' \, B_l = \braket{ 0 | V^\dag \rho' \otimes \mathbb{I}_\mathcal{A} V | 0} =  \tilde \rho
\end{gather}
with the restriction $\tr[\sum_l B_l B^\dag_l \rho'_{eq}] = \gamma$. That is, we start the system in the thermal state of $H'$ at inverse temperature $\beta$ and we let the state of the ancilla be anything, i.e., maximally mixed. We then let the system and ancilla interact and measure the ancilla in state $\ket{0}$, otherwise discard the process (see Fig.~\ref{fig3}). The global unitary operations are redistricted such that ancilla winds up in state $\ket{0}$ with probability $\gamma / d_\mathcal{A}$. The projection of the ancilla onto state $\ket{0}$ guarantees that the ancilla is back in the state it started at the beginning of the forward process.

The process given in Eq.~\eqref{backward} is very general and it encompasses the case where the forward process is unital. The tilde denotes that the process is not trace preserving and therefore the output is not a unit-trace matrix. Since the forward process is trace preserving, the backward process is unital. 
If the forward process is unital then the backward process would preserve trace. The distribution of energy changes in the backwards process is
\begin{align}
\tilde{P}_B(-\Delta \U) =& \sum_{lmn} \braket{E_m | B^\dag_l | E'_n} \braket{E'_n| \rho'_{eq} |E'_n} \braket{E'_n | B_l | E_m}\notag\\ 
& \quad\times \delta(\Delta \U-(E'_n-E_m)),
\end{align}
where we have used $\delta(-x-(y-z))=\delta(x-(z-y))$. For the normalisation of the backwards distribution we find $\int d\Delta \U \,\tilde{P}_B(-\Delta \U) = \gamma$. The backwards distribution is only a proper normalised probability distribution if the forward channel is unital. The Jarzynksi equality holds for the backward process: $\braket{e^{\beta \Delta \U - \beta \Delta \f}} = 1$ (see Appendix for details).

%%%%%%%%%%%%%%%%%%%%%%%%%%%%%%%%%%%%%%%%%%%%%%%%%
%%%%%%%%%%%%%%%%%%%%%%%%%%%%%%%%%%%%%%%%%%%%%%%%%

If the dynamics of the forward process is non-unital then the backward distribution does not constitute a valid probability distribution and one maybe attempted to abandon the framework. However at this point we turn the mathematical manipulations in order to impose consistency at the level of the thermodynamic laws. As a first step, let us simply renormalise the energy distribution for the backward process such that
\begin{gather}
P_B(-\Delta \U) := \frac{\tilde{P}_B(-\Delta \U)}{\gamma}.
\end{gather}
By a simple manipulation of the above expressions we find the Crooks' relation
\begin{gather}
\frac{P_F(\Delta \U)}{P_B(-\Delta \U)} = e^{\beta \Delta \U - \beta \Delta \f -\beta \x}.
\end{gather}
Its reciprocal is obtained by cross-multiplying.

%%%%%%%%%%%%%%%%%%%%%%%%%%%%%%%%%%%%%%%%%%%%%%%%%
%%%%%%%%%%%%%%%%%%%%%%%%%%%%%%%%%%%%%%%%%%%%%%%%%

%*****************************************
%*****************************************
{\bf Excess energy.---}
The first expression is of the same mathematical form as the Crooks relation for work distributions with unitary dynamics with the addition of the $\x$ quantity. One of the profound consequences of the standard Crooks relation is that it shows that the average dissipated work can be related to the relative entropy between the distributions. As we are now dealing with an open dynamics and have not divided the heat and work contributions the concept of dissipated work is ambiguous but it is nevertheless interesting to see if the relative entropy between distributions has any meaning. Taking the log of the first equation and integrating with respect the forward distribution gives us: $K[P_F(\Delta \U)\|P_B(-\Delta \U)] = \beta (\Delta \U - \Delta \f - \x)$ where $K$ is the classical relative entropy or Kullbach-Liebler divergence $K[P_F(\Delta \U)\|P_B(-\Delta \U)]=\int d\Delta \U \, P_F(\Delta \U)\log
({P_F(\Delta \U)}/{P_B(-\Delta \U)}) $. 
Rearranging we get the following interesting relationship for the average internal energy change of an open quantum system
\begin{gather}\label{eq11}
\Delta \U=\beta^{-1} K[P_F(\Delta \U)\|P_B(-\Delta \U)] + \x +\Delta \f.
\end{gather}
This expression constitutes a central result of our paper. The equation is consistent with what is already known. Suppose our map is a unitary operation, then the map is trivially unital and therefor $\x=0$. In addition in the unitary case no heat is exchanged with the bath and the average internal energy change can be associated with the average work $\langle \w \rangle=\Delta \U$, in this case we see that the relative entropy term plays the role of the average dissipated work. For open dynamics that is described by a unital map, again $\x=0$ here we may interpret the relative entropy term as the energetic dissipation to the ancilla. Turning the most general non-unital case it is clear that now the $\x$ is nonzero and is related to both the process and the final equilibrium state of the system in the virtual isothermal process -- most importantly it is a term which can be either negative or positive. Consider the generalised first law given as Eq.~\eqref{first} one can say that the average contribution to the energy change which is not the free energy difference is 
\begin{align}
\braket{\bf E}_{excess}=&\beta^{-1} K[P_F(\Delta \U)\|P_B(-\Delta \U)] + \x \notag\\
=&\braket{\w}_{diss} + \braket{\q}. 
\end{align}
For unital process this term is always positive, non-unital dynamics therefore can lead to an energy excess which can be negative, this is related to the environment being allowed to perform work on the system in this case thus leading to process whereby the von Neumann entropy of the system may decrease (only non-unital channels allow this decrease, a physical example of this is cooling which by definition requires a non-unital channel). We now turn towards the issue of the entropy change. 

%*****************************************
%*****************************************
{\bf Non-equilibrium Helmholtz equation.---}
One of the cornerstones of equilibrium thermodynamics is the Helmholtz equation which states
\begin{gather}
\label{helm}
\U = \f + \beta^{-1} \s,
\end{gather} 
where is $\U$ is the internal energy, $\f$ is the equilibrium free energy and $\s$ is the entropy consistent with Eqs.~\eqref{first} and \eqref{second}. We stress here the fact that the von Neumann entropy and the thermodynamical entropy can only be equated when the system under scrutiny is in thermodynamic equilibrium.

In quantum mechanics the internal energy is given by $\tr[\rho H]$. With a little manipulation we get
\begin{align}
\U =& \tr[\rho H] = -\beta^{-1} \tr\left[ \rho \log \left( e^{-\beta H} \right)\right] \notag\\
=&  -\beta^{-1} \tr\left[ \rho \log \left( e^{-\beta H}/Z \right)\right] -\beta^{-1} \tr\left[ \rho \log (Z)\right] \notag\\
=& -\beta^{-1} \tr\left[ \rho \log \left( \rho_{eq} \right)\right] + \f
\end{align}
where $Z=\tr[e^{-\beta H}]$ and $\f=-\beta^{-1} \tr\left[ \rho \log (Z)\right]$ is the equilibrium free energy. Let us define the first term in the last equation as a non-equilibrium entropy
\begin{align}
\s =& -\tr\left[ \rho \log \left( \rho_{eq} \right)\right] - S_V(\rho) + S_V(\rho) \notag\\
=& S_R (\rho \| \rho_{eq}) + S_V(\rho),
\end{align}
where $S_V(\rho) = -\tr[\rho \log(\rho)]$ is the von Neumann entropy and $S_R(\rho \| \sigma) = -\tr[\rho \log(\sigma) - \rho \log(\rho)]$ is the quantum relative entropy. Note that the relative entropy term goes to zero if and only if the system is in thermal equilibrium and $\s$ is equivalent to the von Neumann entropy. Given the consistency of $\s$ with Eq.~\eqref{first}, Eq.~\eqref{second} and Eq.~\eqref{helm} we suggest that it is the correct entropy to use for a generic non-equilibrium transformation on a thermal state. 

Recalling again the first and second law and using the non-equilibrium Helmholtz equation we have $\Delta\U =  \beta^{-1} \Delta\s + \Delta\f$. Now taking Eq.~\eqref{eq11} we obtain 
\begin{align}
 \Delta \s=& \beta \braket{\w}_{diss} + \beta \braket{\q} \notag\\
=&K[P_F(\Delta \U)\|P_B(-\Delta \U)] + \beta \, \x.
\end{align}
The result is a microscopic expression for the entropy change of a quantum system which only depends on the system state and the 
dynamical map $\Lambda$. This law is the second central result of our work. One maybe tempted to equate the dissipated work with the Kullback-Liebler divergence and $\x$ with the heat exchange. However, this cannot be true since for unital process heat is non-vanishing, but $\x=0$. When the dynamics is unital we see that this entropy change is positive and also that it maybe calculated via the relative entropy between the energy distribution of the process and the corresponding backwards distribution. In the non-unital case we see that the $\x$ term comes into play and may become negative and we get the clear picture that non-unital channels can lead to a entropic decrease in the system. Physical examples of this include cooling, some erasure protocols and spontaneous emission. 

Lastly, it is interesting to note the change of information theoretic (von Neumann) entropy:
\begin{gather}
\Delta S_V = K[P_F(\Delta \U)\|P_B(-\Delta \U)] + \x -S_R (\rho' \| \rho'_{eq}).
\end{gather}
Suppose we have a closed system, then the heat exchange $\braket{\q}=0$ and we can ascribe the change of internal energy to work and hence $P(\Delta \U)=P(\w)$. In addition the map generating the dynamics is unitary and by default unital such that $\x=0$. Furthermore, in this case $\Delta S_V=0$ due to unitarity and we arrive at the known result that $\beta \langle \w\rangle_{diss}=K[P_F(\Delta \U)\|P_B(-\Delta \U)]= S_R (\rho' \| \rho'_{eq})$ \cite{deffner}. 

{\bf Conclusion.---}
In this work we have formulated and analysed fluctuation-like relations on the distribution of energetic changes of an open quantum system undergoing evolution described by a completely positive and trace preserving map. For non-unital channels the backward energy distribution is not normalised. By renormalising the distribution and deriving a relation which is mathematically similar to the Crooks relation we were able to derive interesting expression for the energy and entropic changes. These expressions divide both the energetic and entropic changes into terms which are related to specific details of the microscopic dynamics. We believe that this division maybe useful in order to analyse in detail the energetics of complex quantum systems.

{\bf Acknowledgements.---}
We thank Tony John George Apollaro, Felix Binder, Lucas Celeri, Ross Dorner, Eduardo Mascarenhas, Laura Mazzola, Mauro Paternostro, Roberto Sarthour, Vlatko Vedral, and Sai Vinjanampathy for insightful conversations. We thank the John Templeton Foundation and the Ministry of Education \& the National Research Foundation of Singapore for support. This work was partially supported by the COST Action MP1209.

\begin{widetext}

\appendix

{\bf Appendix.---}
Normalization of the forward process:
\begin{align}
\int d\Delta \U \, P_F(\Delta \U) =& \int d\Delta \U \, \sum_{lmn}  \delta(\Delta \U-(E'_n-E_m))  \braket{E'_n | A_l | E_m} \braket{E_m| \rho_{eq} |E_m} \braket{E_m | A^\dag_l | E'_n}  \notag\\
=& \sum_{lm} \braket{E_m | A^\dag_l \sum_n | E'_n}\braket{E'_n | A_l | E_m} \braket{E_m| \rho_{eq} |E_m} 
= \sum_{m} \braket{E_m | \sum_l A^\dag_l A_l | E_m} \braket{E_m| \rho_{eq} |E_m} \notag\\
=& \sum_{m} \braket{E_m| \rho_{eq} |E_m}  = \mbox{tr}[\rho_{eq}] = 1.
\end{align}

Jarzynski for forward process:
\begin{align}
\braket{e^{-\beta \Delta \U + \beta \Delta \f}} =&
\int d\Delta \U P_F(\Delta \U) \, e^{-\beta \Delta \U + \beta \Delta \f}= \int d\Delta \U \,
\sum_{lmn} \braket{E'_n | A_l | E_m} \braket{E_m| \rho_{eq} |E_m} \braket{E_m | A^\dag_l | E'_n} \delta(\Delta \U-(E'_n-E_m)) e^{-\beta \Delta \U + \beta \Delta \f}\notag\\
=& \sum_{lmn} \braket{E'_n | A_l | E_m} \frac{e^{-\beta E_m}}{Z} \braket{E_m | A^\dag_l | E'_n} e^{-\beta E'_n} e^{\beta \Delta \f}
= \sum_{ln} \braket{E'_n | A_l |\sum_m E_m} \braket{E_m | A^\dag_l | E'_n} \frac{e^{-\beta E'_n}}{Z'} e^{\beta \Delta \f }\frac{Z'}{Z}\notag\\
=& \sum_{ln} \braket{E'_n | A_l A^\dag_l | E'_n} \frac{e^{-\beta E'_n}}{Z'}  
= \mbox{tr} \left[ \sum_{l} A_l A^\dag_l \rho'_{eq} \right] = \gamma = e^{-\beta \x}.
\end{align}

For the normalisation of the backwards distribution we find
\begin{align}
\int d\Delta \U \, P_B(-\Delta \U) =& \int d\Delta \U \, \sum_{lmn} \braket{E_m | B^\dag_l | E'_n} \braket{E'_n| \rho'_{eq} |E'_n} \braket{E'_n | B_l | E_m} \delta(\Delta \U-(E'_n-E_m))\notag\\
=& \sum_{ln} \frac{e^{-\beta E'_n}}{Z_n} \braket{E'_n | B_l | \sum_m E_m} \braket{E_m | B^\dag_l | E'_n}
= \mbox{tr}\left[ \sum_l B_l B^\dag_l \rho'_{eq} \right].
\end{align}

Jarzynski for backward process:
derive a Crooks like relation let us first prove the following 
result:
\begin{align}
\braket{e^{\beta \Delta \U - \beta \Delta \f}} =&
\int d\Delta \U \tilde{P}_B(-\Delta \U) \, e^{\beta \Delta \U - \beta \Delta \f}  
= \int d\Delta \U \sum_{lmn} \braket{E_m | B^\dag_l | E'_n} \braket{E'_n| \rho'_{eq} |E'_n} \braket{E'_n | B_l | E_m} \delta(\Delta \U-(E'_n-E_m)) e^{\beta \Delta \U - \beta \Delta \f}\notag\\
=&\sum_{lmn} \braket{E_m | B^\dag_l | E'_n} \frac{e^{-\beta E'_n}}{Z'} \braket{E'_n | B_l | E_m} e^{\beta E'_n- \beta E_m - \beta \Delta \f}
=\sum_{m} \braket{E_m | \sum_l B^\dag_l | \sum_n E'_n} \braket{E'_n | B_l | E_m} \frac{e^{- \beta E_m}}{Z} e^{-\beta \Delta \f} \frac{Z}{Z'}\notag\\
=&\sum_{lmn} \braket{E_m | E_m} \frac{e^{- \beta E_m}}{Z} = 1.
\end{align}
\end{widetext}
\end{document}